\documentstyle[preprint,prb,aps]{revtex}
\tightenlines
\begin{document}
\draft
\title{Magnetoplasmons in quantum rings}
\author{ A. Emperador$^1$, M. Barranco$^2$\cite{perm}, E. Lipparini$^2$, 
M. Pi$^1$, and Ll. Serra$^{3}$.}
\address{$^1$Departament d'Estructura i Constituents de la Mat\`eria,
Facultat de F\'{\i}sica, \\
Universitat de Barcelona, E-08028 Barcelona, Spain}
\address{$^2$Dipartimento di Fisica, Universit\`a di Trento,
and INFM sezione di Trento, I-38050 Povo, Italy}
\address{$^3$Departament de F\'{\i}sica, Facultat de Ci\`encies,\\
Universitat de les Illes Balears, E-07071 Palma de Mallorca, Spain}
\date{\today}

\maketitle

\begin{abstract}
We have studied the structure and dipole charge density response of 
nanorings as a function of the magnetic 
field using local-spin density functional theory.
Two small rings consisting of 12 and 22 electrons confined by a
positively charged background  are used
to represent the cases of a  narrow and a  wide
ring. The results are qualitatively compared with experimental data
existing on  microrings and on antidots. A  smaller ring containing
5 electrons is also analyzed to allow for a closer comparison with
a recent experiment on a two electron quantum ring.

\end{abstract}
\narrowtext
\section*{}

\section{Introduction}

The study of collective excitations in bounded two-dimensional electron
systems (2dES) is a subject of current interest, especially for the particular
geometry called quantum dot in which a  number of electrons  
is confined into a rather small, almost two-dimensional region produced by present 
available etching technologies, and for the quasi-one-dimensional structures
called quantum wires  (see for example Refs. \onlinecite{Jac98,And98} for a 
comprehensive description of quantum dots and wires). Less effort
has been put in the investigation of these excitations in quantum antidots, 
i.e., the reversed structure of dots 
made in the 2dES\cite{Ker91,Zha92,Lor92,WuZ93,Mik96,Emp98}.       

Recently, magnetoplasmons arising in  ring confining geometry 
have also attracted some interest.
The first experimental studies  concerned 
structures in the micron scale, etched into a 
molecular-beam-epitaxy-grown $\delta$-doped GaAs-Ga$_x$Al$_{1-x}$As
heterostructure, of outer diameter $\sim $ 50 $\mu$m and inner
diameter in the 12-30 $\mu$m range. The observed magnetoplasmon 
resonances\cite{Dah93} bear some of the properties of the
dynamical response of a classical 2dES\cite{Pro92}.
Later on, a hydrodynamic theory based on the Thomas-Fermi-Dirac-von 
Weizs\"acker approximation has been used\cite{Zar96} 
to describe $N$ = 400 electron rings which yields
a  good account of the experimental data after an appropiate
scaling of them at zero magnetic field ($B$).
Plasmon modes in very  narrow rings have been described within 
a Hartree$+$random phase approximation\cite{Hua92}, 
and the charge density response of a dot with a repulsive
impurity in its center has also been worked out\cite{Gud94}.
The optical absorption and inelastic scattering of a two electron
quantum ring of a rather large radius (480 nm) and width (20 nm)
has been discussed in detail\cite{Wen96}, and  single electron
properties of quantum rings with parabolic confinement have been
discused\cite{Chak94}, with the aim of determining
the effect of electron-electron interactions on the energy spectrum and
magnetic moment associated with the persistent current in a quantum 
ring\cite{Mai93}.

The far-infrared (FIR) charge density excitation (CDE) appears to
 depend on the ring width. The  measured CDE's\cite{Dah93} 
 are bundled into a high energy group and a low energy group,
which in contradistinction with the case of dots do not merge 
at $B$ = 0. The low energy peaks  arrange into two distinct 
branches. For narrow (NR) rings,  both have a negative $B$ dispersion,
whereas for broad (BR) rings one branch displayes a positive $B$
 dispersion at small magnetic fields.
The high energy peaks arrange into one (narrow rings) or 
several (broad rings) branches. The high energy branches display
 a negative $B$ dispersion at small magnetic fields. 

The low energy peaks have been explained as edge magnetoplasmons
excited at the inner and outer boundaries of the 
ring\cite{Pro92,Zar96}, 
whereas the high energy peaks are bulk
magnetoplasmons\cite{Zar96}. It is worth to recall that 
in the case of antidots, only one edge magnetoplasmon is
detected whose energy goes to zero with $B$\cite{Ker91,Zha92}.
It thus seems that the observed ring plasmons  exhibit
features of either dots or antidots depending on the ring
widthness and $B$ value.  
    
Very recently, nanorings in InAs-GaAs heterostructures
have been fabricated in the 15-40 nm radius range\cite{Gar97},
and the FIR response has been measured for a two electron 
ring\cite{Lor98}. 
Two sets of peaks appear in the response, as in the 
case of microrings. Depending on the $B$ value, one to 
three main peaks have been detected and arranged into four energy
branches with $B$ dispersions which seem to differ
from the microring systematics\cite{Lor98}.
The two branches starting from the $B$ = 0 high energy peak
are similar to those of quantum dots, 
and according to the analysis of the experimental data presented in
Ref. \onlinecite{Lor98}, 
the two branches corresponding to the low energy peaks seem to display
both a positive $B$ dispersion. It is worth to notice that the
experimental results on microrings cover a low $B$ range (up to
2 T), whereas the ones on nanorings extend up to 14 T, but
no data on the low energy  nanoring peaks
have been recorded below 4 T.  

The studied nanorings present an  elongation in
the [1,-1,0] direction. Likely, it is not distorting much the electrons 
from being circularly distributed. Otherwise, one would have that at
$B$ = 0 the two high energy branches do not merge at all, as they
seem to do. A similar situation, namely, a non circularly symmetric
dot hosting quite a circularly symmetric electronic density 
is also found for few electron quantum dots\cite{Kum90}.  
Besides, during their manufacture nanorings had to be
further covered to complete the necessary layer structure\cite{Lor98}.
All that might result in nontrivial changes with respect the CDE's
of a `clean', circularly symmetric ring, and  it       
calls for a microscopic investigation in which the basic
ingredients for a proper description of such nanostructures
are taken into account and might 
guide the experimental  
analysis as a kind of `reference spectrum'
obtained under controled geometrical conditions.    

We present here three such spectra obtained 
within time-dependent local-spin density functional theory 
(TDLSDFT). The first two correspond to circularly
symmetric nanorings made of 12 and 22 electrons embedded
into a GaAs-Ga$_x$Al$_{1-x}$As heterostructure. Although
the method can handle a smaller number of electrons, the 
possibility of describing the two electron structure\cite{Lor98}
is beyond its reach and for that reason we have renounced 
to it from the start, presenting only results obtained for a 
5 electron nanoring of equal size as a third example.
It is dobtless that the natural evolution of the field
will make it possible a quantitative comparison between
TDLSDT and experimental results still to come. 

This work is organized as follows. We discuss in Sect. II the
results we have obtained for the ring ground states (gs), 
which are the starting point for the study of their
charge density excitations presented in Sec. III.
Finally, the concluding remarks are given in Sec. IV.

\section{The ground state of quantum rings }

We consider a circularly symmetric
  quantum ring made of $N$ electrons moving in the
$z$ = 0 plane where they are confined 
by the  potential $V^+(r)$
created by $N^+$ positive charges 
uniformy distributed between an outer $R_o$ and
inner $R_i$ radius in the presence of a constant magnetic 
field $B$ in the positive $z$ direction.
In the local-spin density approximation (LSDA), the single electron
wave functions are given by the
solution of the Kohn-Sham (KS) equations
\begin{eqnarray}
\left[-{1\over 2}\nabla^2\right. &+& {1\over 2}\omega_c\ell_z +
{1\over 8}\omega_c^2r^2 - V^+(r) \nonumber\\
&+& \left. V^H + V^{xc} +(W^{xc}+{1\over 2}g^*\mu_B B)\sigma_z \right]
\varphi_{\alpha}(r,\theta)=\epsilon_{\alpha}\varphi_{\alpha}(r,\theta)~,
\label{eq1}
\end{eqnarray}
where
$V^H=\int{d\vec{r}\,'\rho(\vec{r}\,')/|\vec{r}-\vec{r}\,'|} $
is the Hartree potential.
$V^{xc}={\partial
{\cal E}_{xc}(\rho,m)/\partial\rho}\vert_{gs}$ and
$W^{xc}={\partial
{\cal E}_{xc}(\rho,m)/\partial m}\vert_{gs}$
are the  variations of the exchange-correlation
energy density ${\cal E}_{xc}(\rho,m)$ in the  local  approximation
taken at the ground state, and $\rho(r)$ and $m(r)$ are the
electron and spin magnetization densities. The exchange-correlation
energy density ${\cal E}_{xc}$ has been  constructed from
the results on the nonpolarized and fully
polarized two dimensional  electron gas\cite{Tan89} 
using the two dimensional von Barth and 
Hedin\cite{vBa72} prescription to interpolate between both regimes.

We have used effective atomic units
($\hbar={e^2/\epsilon} = m = 1$), where $\epsilon$
is the dielectric constant of the semiconductor and $m$ is the electron
effective mass.
In units of the bare electron mass $m_e$ one has $m = m^*m_e$.
In this system of units,
the length unit is the effective Bohr radius $a_0^*=a_0\epsilon/m^*$,
and the energy unit is the effective Hartree
$H^*=H m^*/\epsilon^2$.
For GaAs we have taken $\epsilon$ = 12.4, $m^*$ = 0.067, and
$g^*=-0.44$, which yields
$a_0^*=97.9\, {\rm \AA}$ and $H^*\sim11.9$ meV.
In Eq.\ (\ref{eq1}) $\omega_c =e B/(m c)$ is the cyclotron
frequency
and $\mu_B={e\hbar/(2m_e c)}$ is the Bohr magneton.

As a consequence of circular symmetry the
$\varphi_{\alpha}$'s  are eigenstates of the
orbital angular momentum $\ell_z$, i.e.,
$\varphi_{\alpha}(r,\theta)=u_{n\ell\sigma}(r)e^{-i\ell\theta}$,
with $\ell=0,\pm1,\pm2,\dots$.
The gs electron density is given by
$\rho(r)=\sum_{\alpha} n_{\alpha} |u_{\alpha}(r)|^2$,
while the gs spin magnetization density
is expressed in terms of the
spin of orbital $\alpha$, $\langle\sigma_z\rangle_\alpha$, as
$m(r)=\sum_{\alpha} n_{\alpha}
\langle\sigma_z\rangle_\alpha | u_\alpha(r)|^2$.
The numerical calculations reported in the following have  been
performed at a small but finite temperature  $T \leq$ 0.1 K, and
the KS equations have been solved by integration in $r$ space.
The thermal occupation probabilities $n_{\alpha}$ are determined by the
normalization condition $N=\sum_{\alpha} n_{\alpha} =
\sum_{\alpha} 1/\{1+{\rm exp}[(\epsilon_{\alpha}-\mu)/k_B T]\}$
which fixes the chemical potential $\mu$.



The $V^+(r)$ potential is analytical and can be expressed in terms
of the elliptic ${\bf E}$ and ${\bf K}$ functions\cite{Gra80}:

\begin{eqnarray}
V^+(r)&=&\frac{4 N^+}{\pi (R_o^2 - R_i^2)} 
\label{eq2}
\\
&\times& \left\{ \begin{array}{ll}
\left[ R_o {\bf E}\left(\frac{r}{R_o}\right) - 
R_i {\bf E}\left(\frac{r}{R_i}\right)\right]
\;\;\;\;\;\;\;\; {\rm if} \;  r < R_i & \\
\left[ R_o {\bf E}\left(\frac{r}{R_o}\right) - 
 r {\bf E}\left(\frac{R_i}{r}\right)\right] +
 r \left[ 1 - \left(\frac{R_i}{r}\right)^2\right] 
{\bf K}\left(\frac{R_i}{r}\right)
\;\;\;\;\;\;\;\; {\rm if} \; R_o > r > R_i &\\
r \left\{{\bf E}\left(\frac{R_o}{r}\right) - 
  {\bf E}\left(\frac{R_i}{r}\right)  
+  \left[ 1 - \left(\frac{R_i}{r}\right)^2\right] 
{\bf K}\left(\frac{R_i}{r}\right) -
  \left[ 1 - \left(\frac{R_o}{r}\right)^2\right] 
{\bf K}\left(\frac{R_o}{r}\right)\right\} 
  & \;\;\; {\rm if} \; r > R_o \; .
\end{array}
\right.
\nonumber
\end{eqnarray}

As previously indicated, we have considered two nanorings. The narrow
one has $R_o$ = 100 nm, $R_i$ = 70 nm, $N$ = 12 and $N^+$ = 14, and 
the broad one has $R_o$ = 100 nm, $R_i$ = 37.5 nm, $N$ = 22,
 and $N^+$ = 24. These values have been selected to roughly
have in both rings the same average surface densities as in the
$N$ = 25 quantum dot described in Refs. \onlinecite{Dem90,Lip97,Ser98}, 
as well as the
same outer radius. That would allow to make a comparison
between FIR modes arising in somehow similar
dot and ring geometries. The radii ratio in the broad ring is 
similar to that of Ref. \onlinecite{Lor98}.

Figure \ref{fig1} represents several electron
densities for selected $B$ values as a function of the radial distance
in the case of the NR ring, and Fig. \ref{fig2} in the case of the BR
ring. In the latter case, at $B$ = 0 the central electron density is not  
zero, but it is around two orders of magnitude smaller than its
maximum value. At present, it is unclear to us whether a different
confining potential that  prevents the electrons 
from having a sizeable probability of being inside the ring `hole', as the
parabolic confinement of Ref. \onlinecite{Chak94},
would be more realistic. 

In the case of the NR ring, the electronic density has no structure,
presenting a gaussian-like shape whose width decreases with increasing $B$.
In contradistinction, in the BR ring an
 incipient bulk density region
appears as well as the characteristic `bump' at the edges
clearly visible in dots confined by a disk geometry\cite{Lip97}.

Figures \ref{fig3} and \ref{fig4} represent the single particle (sp) 
energies as a function of the orbital angular momentum $\ell$ 
and different $B$ values. The $N$ = 12 ring becomes fully polarized 
between $B$ = 2 and 3 T, and the $N$ = 22 ring between $B$ = 3 and
4 T (in the $N$ = 25 quantum dot\cite{Lip97} it happens at 
$B \sim 3.6 $ T). It can be seen from the corresponding panels  
in these figures that at $B$ = 0 both rings have a  $z$ 
component of the total spin 
different from zero, $S_z$ = 1. Since the $N$ = 10 and 20 rings are 
close shell systems, this means that Hund's
first rule is obeyed by these small rings, as it is in small 
dots\cite{Tar96}.

The sp energies are arranged into bands which are bent upwards at
both ends not only at low $B$. 
This is a peculiarity of the ring geometry, which
bears simultaneously the characteristics of dot and antidot 
bands, the former ones bending upwards at high $\ell$, and the 
later ones at small $\ell$\cite{Emp98b}. The existence of
two  bendings when a magnetic field is applied
is the microscopic origin of the two edge 
magnetoplasmons, as we shall discuss in the next Section.  

When the ring becomes fully polarized,  
increasing  $B$ further produces the
displacement as a whole of the set of occupied sp levels 
to higher $\ell$'s. We have found that this is the mechanism  
rings have to keep 
its total orbital angular momentum $L_z$  increasing with $B$.
That can be seen for example,
in the high $B$ panels corresponding to  the NR ring (see also Fig. 
\ref{fig11}). We have plotted in Fig.
\ref{fig5} the evolution  of $L_z$ and $2 S_z$ with $B$ for the 
NR ring, and in Fig. \ref{fig6} for the BR ring. 

The shifting upwards in $\ell$ of the whole sp spectrum with
increasing $B$ is a distinct characteristic of rings 
that deserves further investigation.
In quantum dots, the  stability region in the $N - B$ phase plane 
of the fully polarized
configuration, called maximum density droplet (MDD) state,
built from sp orbitals having $\ell$ = 0,1,2 ... 
$N-1$\cite{Mac93,Cha94} is limited from the left by a line
$B_f$ representing, for a given number number of electrons,
the magnetic field at which $2 S_z = N$, and from the right by a 
 line $B_r$ at which edge reconstruction 
starts\cite{Yan93,Kle95,Fer97}. This is a rather narrow region, a few 
 tenths of tesla wide\cite{Fer97} because after fully polarization
the magnetic field is very effective in promoting electrons from
high  to higher $\ell$ sp levels, reconstructing the dot edge. 
In rings this is quite not so because the existence of an electron depletion
at the center and the consequent upwards bending of the sp bands allows for
an alternative mechanism to keep increasing $L_z$ while retaining the
simplicity of the gs wave function, namelly, a Slater determinant
made of the lowest possible $\ell$ sp states from a minimum $\ell_m$ to a 
maximum $\ell_M$ such that $N = \ell_M - \ell_m +1$. Taking as an example 
the situation of the NR ring, at $B$ = 11 T we have found that $\ell_m$ =
54 and $\ell_M$ = 65. It might
well happen that for quantum rings, no equivalent of a kind of
edge reconstruction mechanism exists, 
but addressing this point
is beyond the capabilities of the density functional we are using.   

\section{Charge density excitations of a quantum ring.}

\subsection{Longitudinal response within TDLSDT}

Once the gs has been obtained, we
determine the induced densities originated by an external field
employing linear-response theory.
Following Refs. \onlinecite{Wil83,Raj78}, we can write
the variation $\delta\rho_{\sigma}$ induced in the spin density
$\rho_{\sigma}$  ($\sigma\equiv\uparrow,\downarrow$) by an external
spin-dependent field $F$, whose non-temporal dependence we denote as
$F=\sum_{\sigma}f_{\sigma}(\vec{r})\,|\sigma\rangle\langle\sigma|$:
\begin{equation}
\delta\rho_{\sigma}(\vec{r},\omega) =
\sum_{\sigma'}\int d\vec{r}\,'\chi_{\sigma\sigma'}
(\vec{r},\vec{r}\,';\omega)f_{\sigma'}(\vec{r}\,')\; ,
\label{eq3}
\end{equation}
where
$\chi_{\sigma\sigma'}$ is the spin-density correlation function.
In this limit, the frequency $\omega$ corresponds to the
harmonic time dependence of the external field $F$ and   of
the induced $\delta\rho_\sigma$. Eq.\ (\ref{eq3}) is a 2$\times$2
matrix equation in the two-component Pauli space.
In longitudinal response theory, $F$ is diagonal in this space,
and we write its diagonal components as a vector
$F\equiv\left(
\begin{array}{c} f_\uparrow\\ f_\downarrow\end{array}
\right) $.
 For the dipole operator we then have\cite{note}

\begin{equation}
D_{\rho} \equiv   \left(\begin{array}{c} x\\ x\end{array} \right)
\,\,\,\,\,\,{\rm and}\,\,\,
D_{m} \equiv \left(\begin{array}{c} x\\ -x\end{array} \right) \,\,\, ,
\label{eq4}
\end{equation}
where the field $D_m$ will cause longitudinal spin excitations
not quite studied here because of the lack of experimental 
information on them, but introduced at this point for the sake of
clearness. 

 TDLSDT assumes that electrons respond as  free particles
to the perturbing effective field, which consists of the external
plus the induced field arising from  the changes produced
by the perturbation in the gs mean field. This
condition defines the TDLSDT correlation function
$\chi_{\sigma\sigma'}$ in terms of the
free particle spin-density correlation function
$\chi^{(0)}_{\sigma\sigma'}$ through a
Dyson-type integral equation:
\begin{eqnarray}
\chi_{\sigma\sigma'}(\vec{r},\vec{r}\,';\omega) &=&
\chi^{(0)}_{\sigma\sigma'}(\vec{r},\vec{r}\,';\omega)\nonumber\\
&+&
\sum_{\sigma_1\sigma_2}\int d\vec{r}_1d\vec{r}_2\,
\chi^{(0)}_{\sigma\sigma_1}(\vec{r},\vec{r}_1;\omega)
K_{\sigma_1\sigma_2}(\vec{r}_1,\vec{r}_2)
\chi_{\sigma_2\sigma'}(\vec{r}_2,\vec{r}\,';\omega) \,\,\, .
\label{eq5}
\end{eqnarray}
The free particle spin-correlation function at finite
temperature  is obtained
from the KS sp wave functions, energies and occupation
probabilities:
\begin{equation}
\chi^{(0)}_{\sigma\sigma'}(\vec{r},\vec{r}\,',\omega)=
-\delta_{\sigma,\sigma'}
\sum_{\alpha\beta}\varphi^*_{\alpha}(\vec{r}\,)
\varphi_{\beta}(\vec{r}\,){n_{\alpha}-n_{\beta}\over
\epsilon_{\alpha}-\epsilon_{\beta} +\omega +i\eta}
\varphi_{\beta}^*(\vec{r}\,')
\varphi_{\alpha}(\vec{r}\,')  \,\,\, .
\label{eq6}
\end{equation}
The label $\alpha$ ($\beta$) refers to a
sp level with spin $\sigma$ ($\sigma'$) and occupation
probability $n_{\alpha}$ ($n_{\beta}$).
To simplify the analysis of the results, we have
added a small but finite imaginary part $\eta$ to the energy $\omega$.
This will make an average of the strength function by transforming
the $\delta$-peaks into Lorentzians of width $2\eta$.

The kernel $K_{\sigma\sigma'}(\vec{r},\vec{r}\,')$
is the residual two-body interaction

\begin{equation}
K_{\sigma\sigma'}(\vec{r}_1,\vec{r}_2) =
{1\over | \vec{r}_1-\vec{r}_2 |} +
\left.{\partial^2{\cal E}_{xc}(\rho,m)
\over\partial\rho_{\sigma}\partial\rho_{\sigma'}}
\right\vert_{gs}\delta(\vec{r}_1-\vec{r}_2)\; ,
\label{eq7}
\end{equation}
where
\begin{eqnarray}
\left.
{\partial^2{\cal E}_{xc}\over
\partial\rho_{\sigma}\partial\rho_{\sigma'}} \right\vert_{gs}
&=&
\left.
{\partial^2{\cal E}_{xc}\over\partial\rho^2}
\right\vert_{gs} +
(\eta_{\sigma}+\eta_{\sigma'})\,
\left.{\partial^2{\cal E}_{xc}\over\partial\rho\,\partial
m}\right\vert_{gs} +
\eta_{\sigma}\eta_{\sigma'}\,
\left.{\partial^2{\cal E}_{xc}\over\partial
m^2}\right\vert_{gs} \nonumber\\
&\equiv&
K(r) +
(\eta_{\sigma}+\eta_{\sigma'})\, L(r) +
\eta_{\sigma}\eta_{\sigma'}\; I(r) \;,
\label{eq8}
\end{eqnarray}
with $\eta_{\uparrow}=1, \eta_{\downarrow}=-1$. The last expression
is the definition of the $K$, $L$, and $I$ functions.

When the system is not polarized, there are only two independent
correlation
functions. These are $\chi_{\rho\rho}$ and $\chi_{mm}$ describing,
respectively,
the density response to $D_{\rho}$ and the spin
response to $D_{m}$. They are given by
\begin{eqnarray}
\chi_{\rho\rho}&=&\chi_{\uparrow\uparrow}+\chi_{\downarrow\downarrow}+
\chi_{\uparrow\downarrow}+\chi_{\downarrow\uparrow} \nonumber\\
\chi_{mm}&=&\chi_{\uparrow\uparrow}+\chi_{\downarrow\downarrow}-
\chi_{\uparrow\downarrow}-\chi_{\downarrow\uparrow}  \,\,\,   ,
\label{eq9}
\end{eqnarray}
and the four equations (\ref{eq5}) reduce to two uncoupled equations
for $\chi_{\rho\rho}$ and $\chi_{mm}$ whose kernels are
given by $1/r_{12} + K \delta(r_{12})$ and $I\delta(r_{12})$,
respectively,  and the free particle correlation
function $\chi^{(0)}=\chi^{(0)}_{\uparrow\uparrow}+
\chi^{(0)}_{\downarrow\downarrow}=
2\chi^{(0)}_{\uparrow\uparrow}$ is the same in both channels because
$\chi^{(0)}_{\uparrow\uparrow} = \chi^{(0)}_{\downarrow\downarrow}$.
This constitutes the paramagnetic limit of the longitudinal response
with uncoupled density and spin channels\cite{Ser97}, in which the
residual interaction consists of a Coulomb direct plus an
exchange-correlation terms in one case,
and only of an exchange-correlation term in the other.

When the system is polarized one no longer has
$\chi^{(0)}_{\uparrow\uparrow} = \chi^{(0)}_{\downarrow\downarrow}$,
and there are two more independent correlation functions
\begin{eqnarray}
\chi_{\rho m}&=&\chi_{\uparrow\uparrow}-\chi_{\downarrow\downarrow}-
\chi_{\uparrow\downarrow}+\chi_{\downarrow\uparrow} \nonumber\\
\chi_{m \rho}&=&\chi_{\uparrow\uparrow}-\chi_{\downarrow\downarrow}+
\chi_{\uparrow\downarrow}-\chi_{\downarrow\uparrow}\; ,
\label{eq10}
\end{eqnarray}
which produce the density response to  $D_{m}$ and the
spin response to $D_{\rho}$, respectively.
Since we are interested only in the charge density response, from now 
on we shall restrict ourselves to the discussion of the 
electron response
to $D_{\rho}$, apart form presenting as an example  how 
the longitudinal spin response looks like in two selected cases. 
We refer the reader to Ref. \onlinecite{Ser98} for a thorough discussion of
the longitudinal response in quantum dots, of direct applicability to
quantum rings.

Equations (\ref{eq5}) have been solved as a generalized matrix
equation in coordinate space after
performing an angular decomposition of  $\chi_{\sigma\sigma'}$ and
$K_{\sigma\sigma'}$ of the kind
\begin{equation}
K_{\sigma\sigma'}(\vec{r},\vec{r}\,')=
\sum_{\ell}K^{(\ell)}_{\sigma\sigma'}(r,r') \,
e^{\,\imath \ell(\theta -\theta')}
\; .
\label{eq11}
\end{equation}
Only modes with $\ell=\pm 1$ couple
to the external  dipole field $D_\rho$.
This can be readily seen performing the
angular integral in Eq. (\ref{eq3}). In practice, we have
considered the
multipole expansion of the external field, using the
dipole vectors
\begin{equation}
D_\rho^{(\pm 1)} = \frac{1}{2} r e^{\pm \imath \theta}
\left(\begin{array}{c} 1\\ 1\end{array}\right)
\,\,\, .
\label{eq12}
\end{equation}
For a polarized system having a non zero magnetization in the gs,
 the $\ell=\pm 1$ modes are not
degenerate and may give rise to two excitation branches with
$\Delta L_z=\pm 1$, where $L_z$ is the gs orbital
angular momentum.

The charge density response function  for the dipole field
has been obtained from the
$\ell=\pm1$ components of the correlation functions
$\chi_{\rho\rho}^{(\pm 1)}(r,r';\omega)$  as:
\begin{eqnarray}
\alpha_{\rho\rho}(\omega)&=& \pi^2\int d r_1\, d r_2 \,r_1^2\, r_2^2\,
(\chi_{\rho\rho}^{(+1)}(r_1,r_2;\omega)+
\chi_{\rho\rho}^{(-1)}(r_1,r_2;\omega)) \nonumber\\
&\equiv&
\alpha_{\rho\rho}^{(+1)}(\omega) + \alpha_{\rho\rho}^{(-1)}(\omega)
\label{eq13}
\end{eqnarray}
Its imaginary part is related to the
strength function as $S_{\rho\rho}(\omega)=
{1\over\pi} {\rm Im}[\alpha_{\rho\rho}(\omega)]$.

To check the numerical accuracy of the calculations we have used
the f-sum rule for the dipole operator, which can be
expressed in terms of gs quantities\cite{Lip89}:
\begin{equation}
m^{(\rho\rho)}_1 = \int S_{\rho\rho}(\omega)\,\omega\, d\omega
={1\over2}\langle 0|[x,[H,x]] | 0\rangle ={N\over2}
\,\, .
\label{eq15}
\end{equation}
We have checked that in our calculations  the f-sum rule is fulfilled
within a  95 \% or better.

\subsection{Numerical results}

Figures \ref{fig7} and \ref{fig8}  show the charge density
strength function for the $N$ = 12 and the $N$ = 22 
electron rings,
respectively, and several  $B$ values. The plus or
minus sign close to the more intense peaks indicates 
that they are originated  either by $D_{\rho}^{(+1)}$ 
or by $D_{\rho}^{(-1)}$. Obviously, at zero magnetic field, 
(+) and ($-$) excitations are degenerated.

The CDE's displayed in these figures are easier to understand
starting from the high $B$ results and having in mind the
sp levels drawn in Figs. \ref{fig3} and \ref{fig4}. First
notice that
as in dots, the  (+) low energy  modes are intraband  CDE's 
from the outer ring  boundary.
The ($-$) low energy modes are intraband CDE's  of the inner ring boundary, 
obviously absent in dots. However, they are  the only edge modes in
antidots\cite{Ker91,Zha92}. 
The (+) modes arise
when the dipole field changes  by +1 the total $L_z$ of the ring, and the
($-$) ones when this change is $-1$. Figs. \ref{fig3} and \ref{fig4}
show than indeed, both kind of {\em edge} modes are possible
in rings. 

The higher energy peaks are bulk modes  arising from interband 
transitions. 
At moderate $B$ values, both positive and negative high energy peaks are
present in the strength, but at high $B$ values only modes excited by
$D_{\rho}^{(-1)}$ have an appreciable intensity: as in the dot case, the
(+) low energy edge mode is taking all the strength corresponding to the
$D^{(+1)}_{\rho}$ operator.

Fig. \ref{fig3}  shows that for some $B$ values, 
the sp energies are distributed following a 
very symmetric pattern as a function of $\ell$. This is the reason why
sometimes (+) and $(-)$ edge modes are nearly degenerated.
Their splitting is not regular as a function of $B$, 
indicating a kind of `shell structure'
effect that only a microscopic model can reveal. Still, the gross
features of the three energy branches displayed in Fig. \ref{fig7} is 
very similar to that of narrow microrings\cite{Dah93,Zar96}: 
two low energy
edge modes with a negative $B$ dispersion, and a high energy mode which
at low $B$ has a negative dispersion, and eventually a positive $B$
dispersion at high magnetic fields.  

At zero magnetic field, or generally speaking, at low $B$, the CDE's
 are delocalized 
as in the case of quantum dots. Notice for instance that
(+) and ($-$) excitations are degenerated, and it has no sense to
associate any of them to excitations coming from the inner or
outer ring boundary. Besides, in the case of BR rings the low $B$
strength in rather fragmented, rendering more complex  the analysis.
It is worth to recall that a similar fragmentation occurs in the case 
of dots if one uses a positively charged disk to model the confining
potential\cite{Ser98}. Still, two quite distinct structures, one
at high and another at low excitation energies are present at 
$B$ = 0  in the case of rings, whose origin can be traced back from
the results at high $B$, and one may associate the low energy peaks 
to intraband and the high energy peaks to interband transitions.   

In the BR ring case, 
the $B$ dispersion of the ($-$) edge mode is firstly positive, 
reaches a maximum at around $B$ = 1 T, and then becomes negative.
The high energy peaks with appreciable strength are now only ($-$)
modes. Again, these features are those displayed by  broad microrings.

For quantum dots and rings of similar size and electron number, one 
expects that the energy of the $B$ = 0 mode is lower for the ring than 
for the dot. Actually, this is a experimental fact\cite{Lor98} that
we can qualitatively explained using a sum rule method.
We have found that at $B$ = 0 the average frequency of the dipole mode 
can be written  as\cite{Lip97}

\begin{equation}
\Omega^2 = \frac{1}{2 N} \int d\,{\vec r} \Delta V^+ (r) \rho(r) 
\,\,\,\, .
\label{eq100}
\end{equation}
Taking a parabola $\omega_0^2 r^2/2$ as confining potential $V^+(r)$ for a dot,
and  $\omega_0^2 (r - R)^2/2$ for a ring of mean radius $R$ having the same
number of electrons, one can easily
check that $\Omega = \omega_0$ for the dot, and $\Omega \sim \omega_0 /
\sqrt{2}$ for a narrow ring, or for a ring  broad enough so that the 
electronic density can be considered as being constant.  
  
The FIR response of BR rings have also some  features in common with
antidots, which we recall that at $B \neq 0$ basically comes
from the $D_{\rho}^{(-1)}$ component of the dipole operator. 
One is the $B$ dispersion of the inner edge mode. The other one is the
transfer of stregth from the low to the high energy ($-$) 
peak\cite{Ker91,Zha92}. 

In all cases we have studied,  CDE's emerge as collective peaks.
The residual electron-hole (e-h) interaction shifts CDE's
to higher energies from the sp excitations (SPE) which constitute 
the free response (see Fig. \ref{fig10} below).  In the
longitudinal spin case, the residual interaction is attractive but
weak, as it is only due to the exchange-correlation potential. As an
example, we show in Fig. \ref{fig9} the three responses for the NR and
BR rings at $B$ = 1 T.
 
Finally, we discuss the results we have obtained for a nanoring
more similar to that experimentally studied\cite{Lor98}.
In this case, $R_o$ = 40 nm, $R_i$ = 15 nm, and $N = N^+$ = 5.
Figure \ref{fig10} shows the charge density strength function at 
several $B$ values, and Fig. \ref{fig11} the sp energy levels. 

Basically, the results are qualitatively similar to those of the
broad nanoring already discussed (they have the same $R_o/R_i$
ratio). When a magnetic field is applied, it
can be clearly seen the transfer of strength between the
 edge and  bulk ($-$) branches as $B$ increases, quite 
similar to the antidot case as we have already 
pointed out. The transfer is possible because both branches
have the same polarization. The coupling 
is very inefficient in narrow rings,  and the ($-$) high and low energy 
peaks keep their own strength. This is the situation 
displayed in Fig. \ref{fig7} for the $N$ = 12 ring. 

It is worth to notice the  evolution with $B$ of the ($-$)
edge mode, which is a rather high energy mode with  a positive $B$ 
dispersion from $B$ = 1 to 4 T, and whose
 energy abrupty falls  between 4.5 and 5 T. This 
decreasing is due to a change in the occupied sp levels which 
illustrates the relevance of shell effects especially in the case
of a small number of electrons. A look at the panels corresponding
to $B$ = 2 and 5 T in Fig. \ref{fig11} explains the effect. It can
be  seen how asymmetrically are distributed the sp levels, 
with a much large  energy difference for the
e-h pairs contributing to the edge excitation of the inner 
ring boundary than for those building the edge excitation of the
outher boundary. This explains the large energy of the ($-$)
edge excitation up to $B \sim $ 4.5 T. Of course, this is a 
qualitative argument since the residual e-h interaction has a
sizeable effect in the charge density channel. On the contrary,
and $B$ = 5 T and above, the sp levels are distributed more
symmetrically, the e-h energy differences are smaller and the
(+) and ($-$) edge modes follow the BR ring systematics.   

\section{Summary and outlook}

In this work we have studied in some detail CDE's in quantum rings.
We have confirmed the expectations put forward by Dahl et 
al\cite{Dah93} that plasmon resonances in quantum rings are dominated
to a large extent by geometric effects, although  shell effects
may cause, in the case of few electron nanorings, effects that 
cannot be systematized.  Apart form an
example, we have restricted our analysis to  CDE's. It would be as simple
to describe SDE's and SPE's  within TDLSDT, much along the case of
quantum dots\cite{Ser98}, if experimental information becomes
available.

Our work  complements the theoretical description
of microrings made by Zaremba\cite{Zar96}.
Even if a kind of characteristic pattern can be established for narrow
or broad nanorings, this confining geometry allows to 
 study much richer spectra than in dots or antidots. It might then offer
the possibility of testing theoretical descriptions that
are equally well describing  plasmon modes in quantum dots, even
if their complexity is quite different. 

The lack of experimental results for nanorings hosting several 
electrons has not allowed us to make a quantitative comparison
of  our calculations with experiments. A qualitative
comparison between the calculated $N$ = 5 and the measured $N$ = 2
FIR spectrum\cite{Lor98} is inconclusive. 
To unambiguously arrange the  peaks into branches and
disentangle the $B$ dispersion of the plasmon modes, it would
be essential to experimentally   assign the polarization 
state to the main energy peaks.  This
has been paramount in the analysis of the theoretical FIR response,
which otherwise would have not allowed us to distinguish between
peak fragmentation and different plasmon branches in some cases. 
Alternatively, calculations for rings with as many electrons as in
the experiments might guide to  distribute the experimental 
data into branches. TDLSDT may be a useful tool for doing so in
nanorings with a few more electrons than those studied so far.
Other more microscopic methods  \cite{Wen96} are better suited 
for a two electron ring provided the geometry is adjusted to the 
experimental situation. 

Finally, we have also determined that Hund's first rule is 
fulfilled in the quantum rings we have studied, and have 
elucidated a possible mechanism by which 
a fully polarized quantum ring may have a rather simple gs 
structure in a wide range of magnetic fields.    

\acknowledgements

We are most indebted to Dr. Axel Lorke for a very useful correspondance.
This work has been performed under Grant Nos. PB95-1249 and
PB95-0492 from CICYT, and 1998SGR00011 from Generalitat of
Catalunya. A.E. and M.B. (Ref. PR1997-0174) acknowledge support
from the DGES (Spain).

\begin{figure}
\caption{Electronic densities (10$^{10}$ cm$^{-2}$)  as a function of
the radial distance (nm) for the narrow ring.}
\label{fig1}
\end{figure}
\begin{figure}
\caption{Same as Fig. 1 for the broad ring.}
\label{fig2}
\end{figure}
\begin{figure}
\caption{Single particle energies as a function of orbital angular
momentum $\ell$ corresponding to the narrow ring.
The horizontal lines represent the electron chemical potential. 
The full, upright triangles represent $\sigma = \uparrow$ bands,
and the empty, downright triangles represent $\sigma = \downarrow$
bands.}
\label{fig3}
\end{figure}
\begin{figure}
\caption{Same as Fig. 3 for the broad ring.}
\label{fig4}
\end{figure}
\begin{figure}
\caption{Total orbital and spin angular momenta
$L_z$ and $2 S_z$ as a function of $B$ for the narrow ring.
The dashed lines are drawn to guide the eye.}
\label{fig5}
\end{figure}
\begin{figure}
\caption{Same as Fig. 5 for the broad ring.}
\label{fig6}
\end{figure}
\begin{figure}
\caption{Strength function (arbitrary units)
 as a function of the excitation energy (meV) for the narrow ring 
at several $B$ values.
The arrows indicate the value of the cyclotron frequency. The ($-$) or
(+) symbol close to the more intense peaks denotes the 
character of the  dipole polarization.} 
\label{fig7}
\end{figure}
\begin{figure}
\caption{Same as Fig. 7 for the broad ring.}
\label{fig8}
\end{figure}
\begin{figure}
\caption{Strength function (arbitrary units) at $B$ = 1 T
 as a function of the excitation energy (meV) for the narrow ring 
(top panel) and the broad ring (bottom panel). The solid line is
the charge density response, the dotted line the longitudinal spin 
density response, and the dashed line the free electron response.}
\label{fig9}
\end{figure}
\begin{figure}
\caption{Same as Fig. 7 for the $N = N^+ = 5$ nanoring
with $R_o$ = 40 nm and $R_i$ = 15 nm. The free strength function 
is also plotted (thin lines).}
\label{fig10}
\end{figure}
\begin{figure}
\caption{Same as Fig. 3 for the $N = N^+ = 5$ nanoring
with $R_o$ = 40 nm and $R_i$ = 15 nm.}
\label{fig11}
\end{figure}


\begin{references}
\bibitem{note} In one-electron operators such as $F$, $D_{\rho}$ and 
$D_m$, a sum on the electron coordinates is implicit.
\end{references}
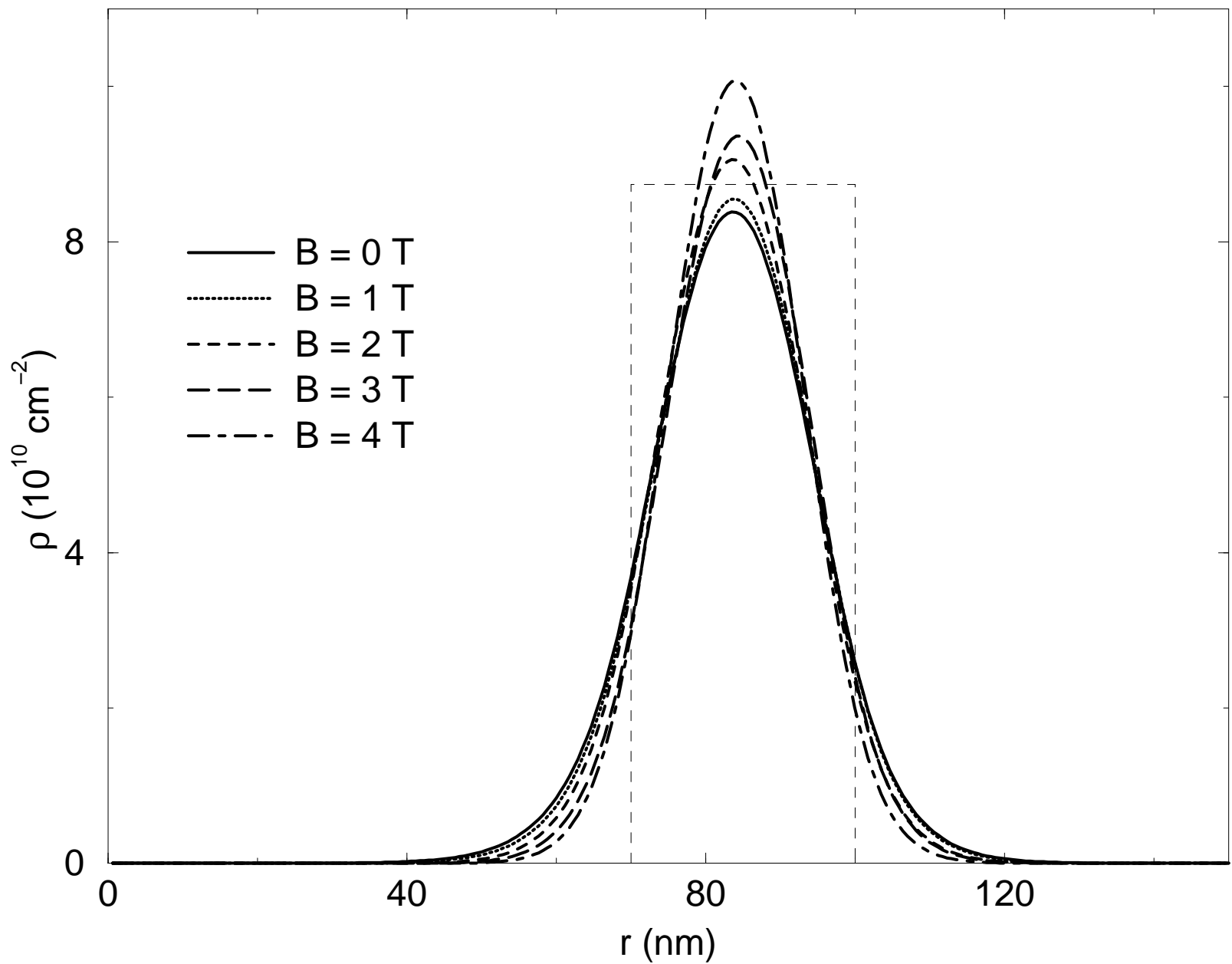
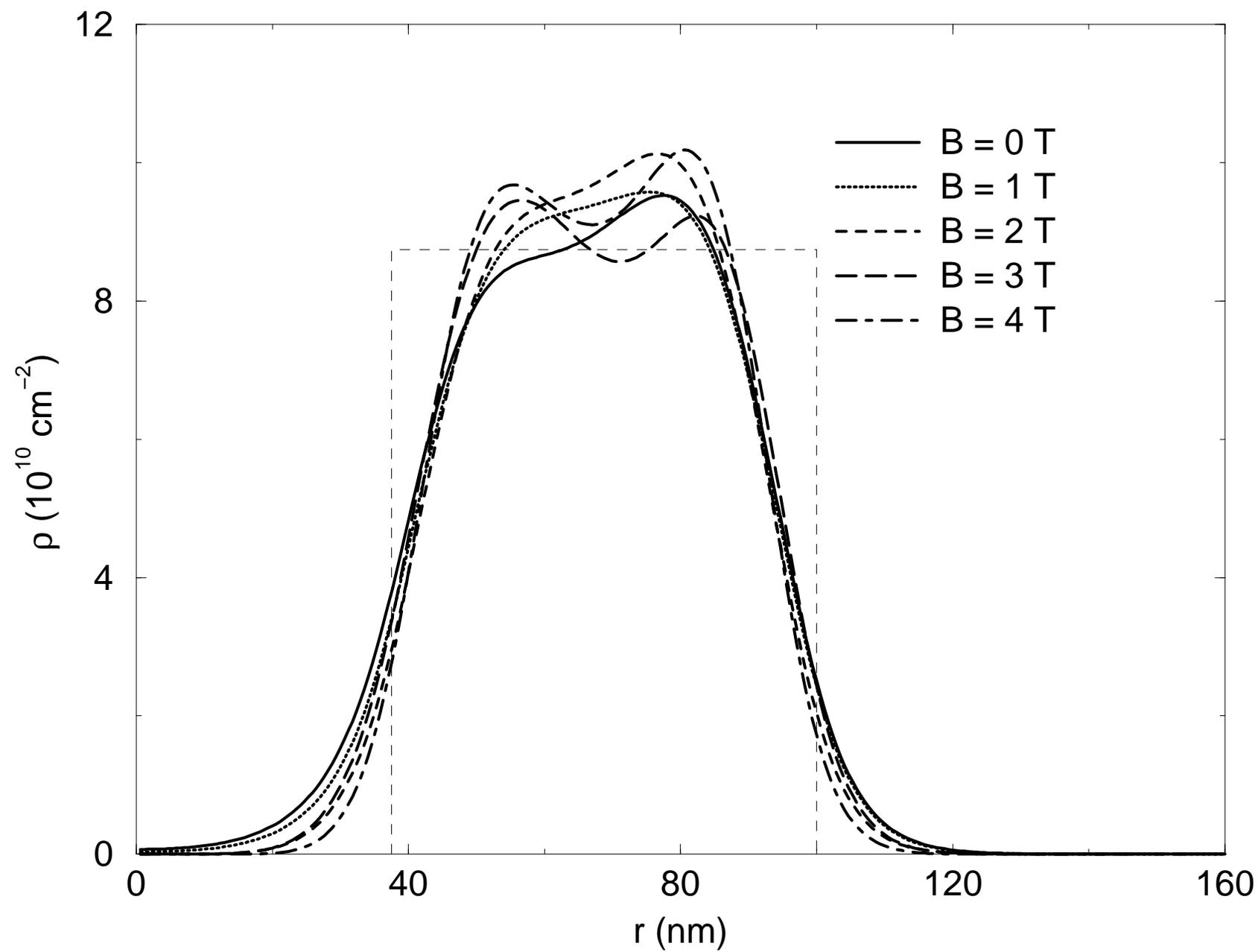
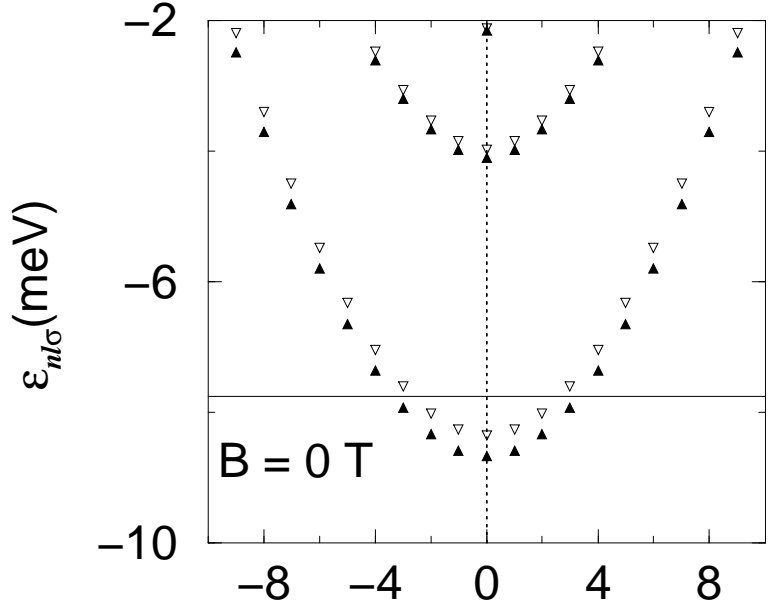
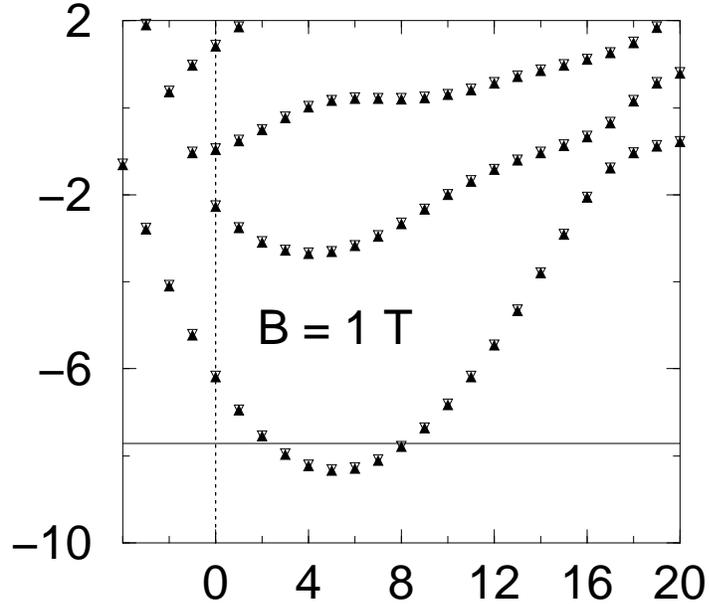
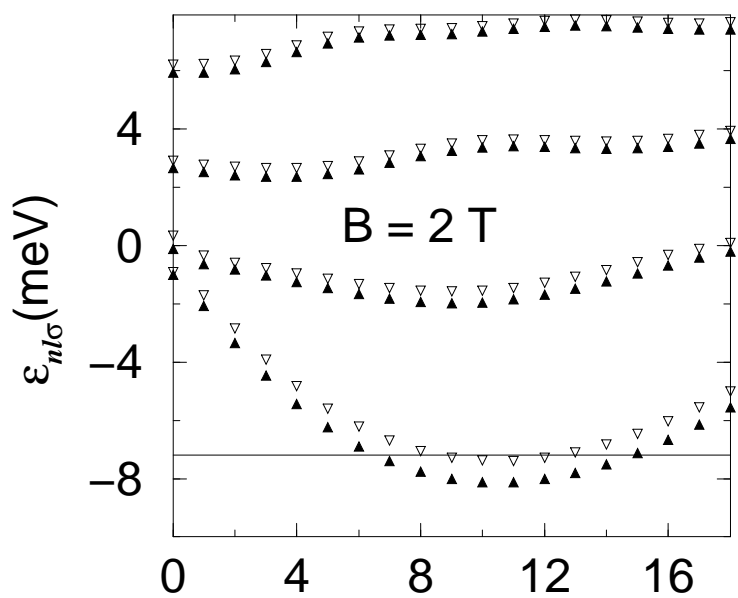
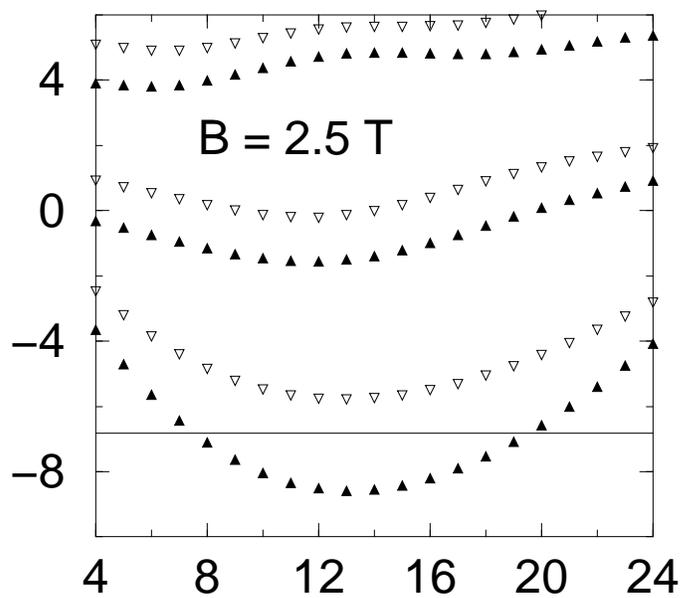
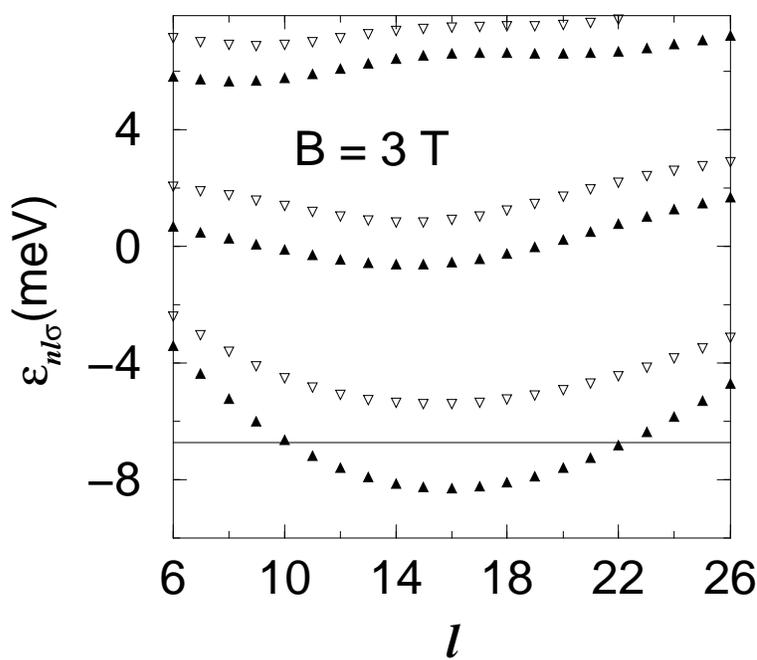
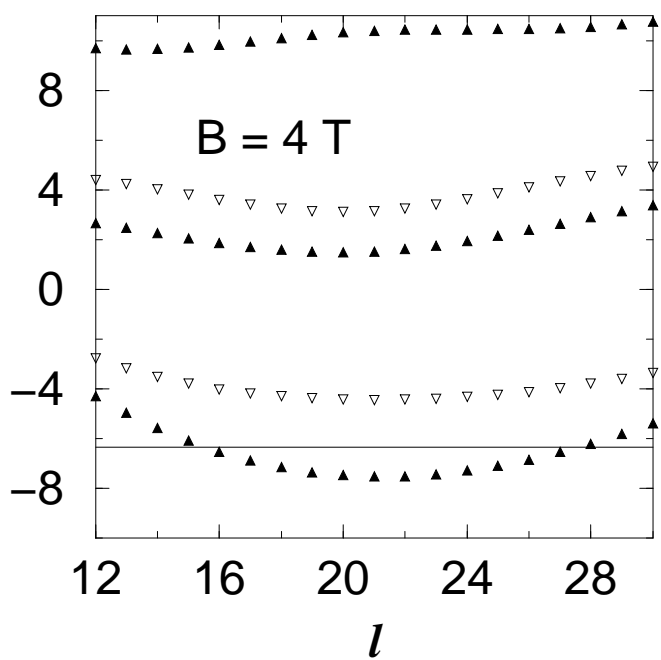
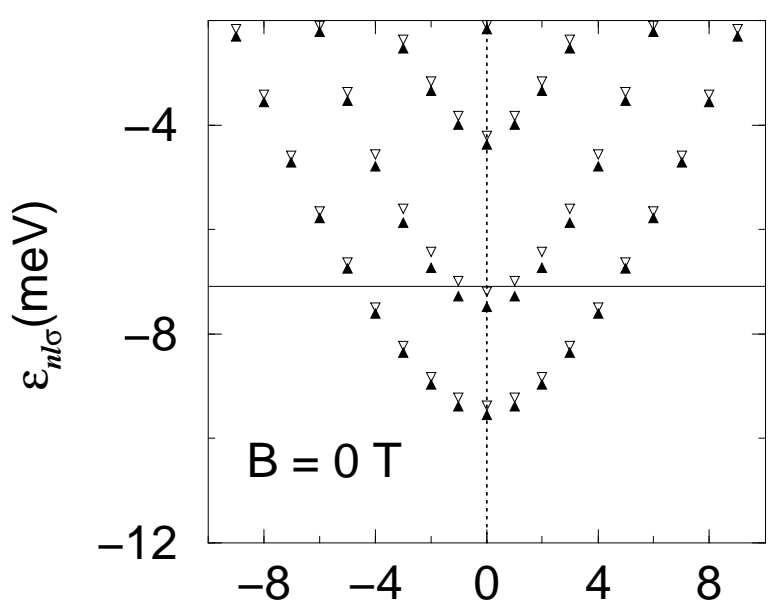
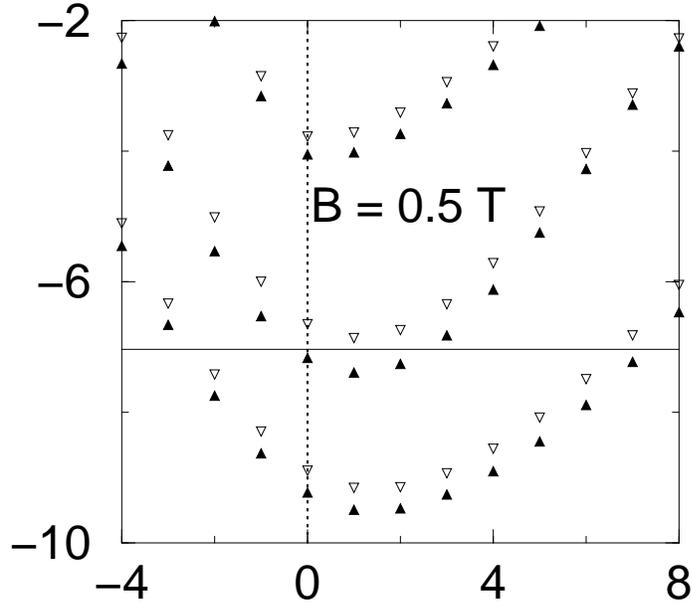
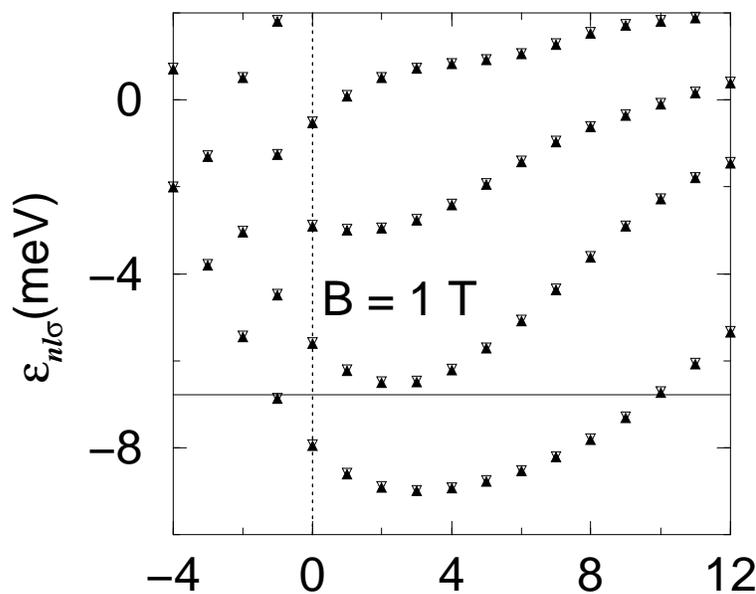
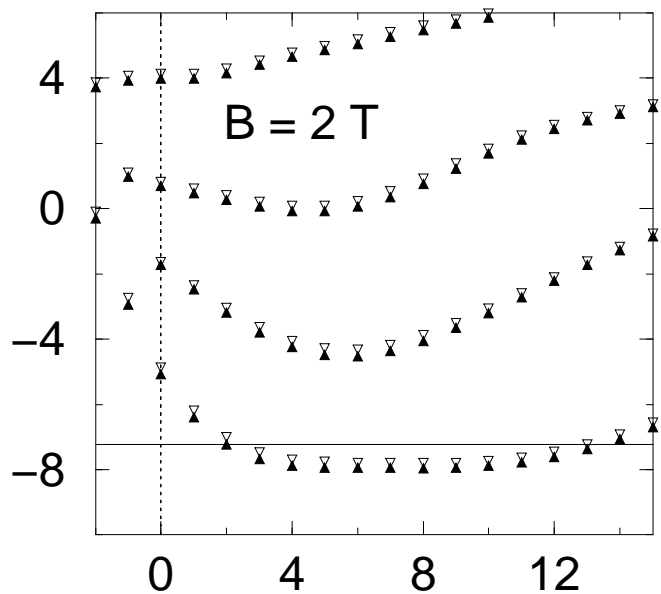
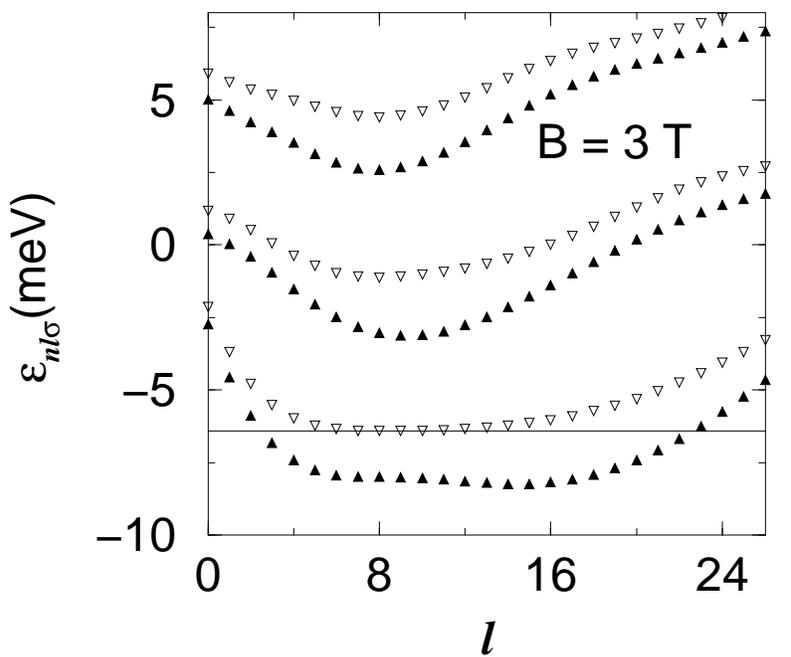
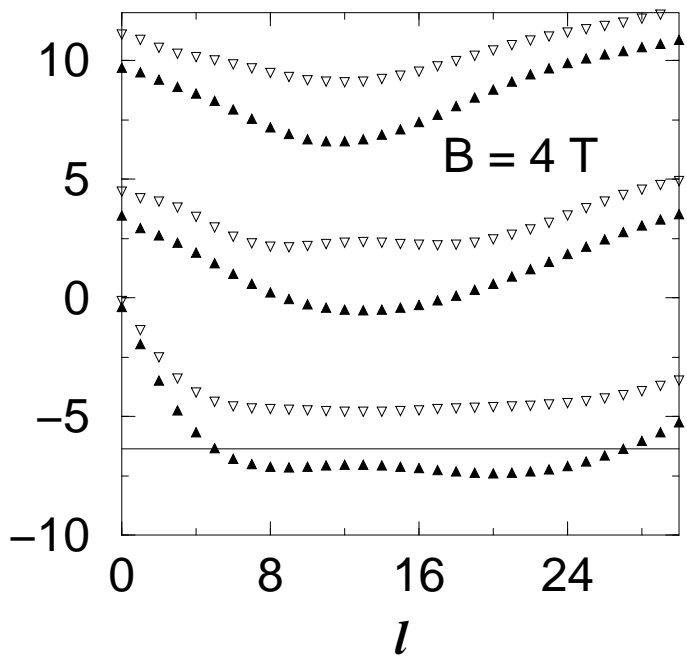
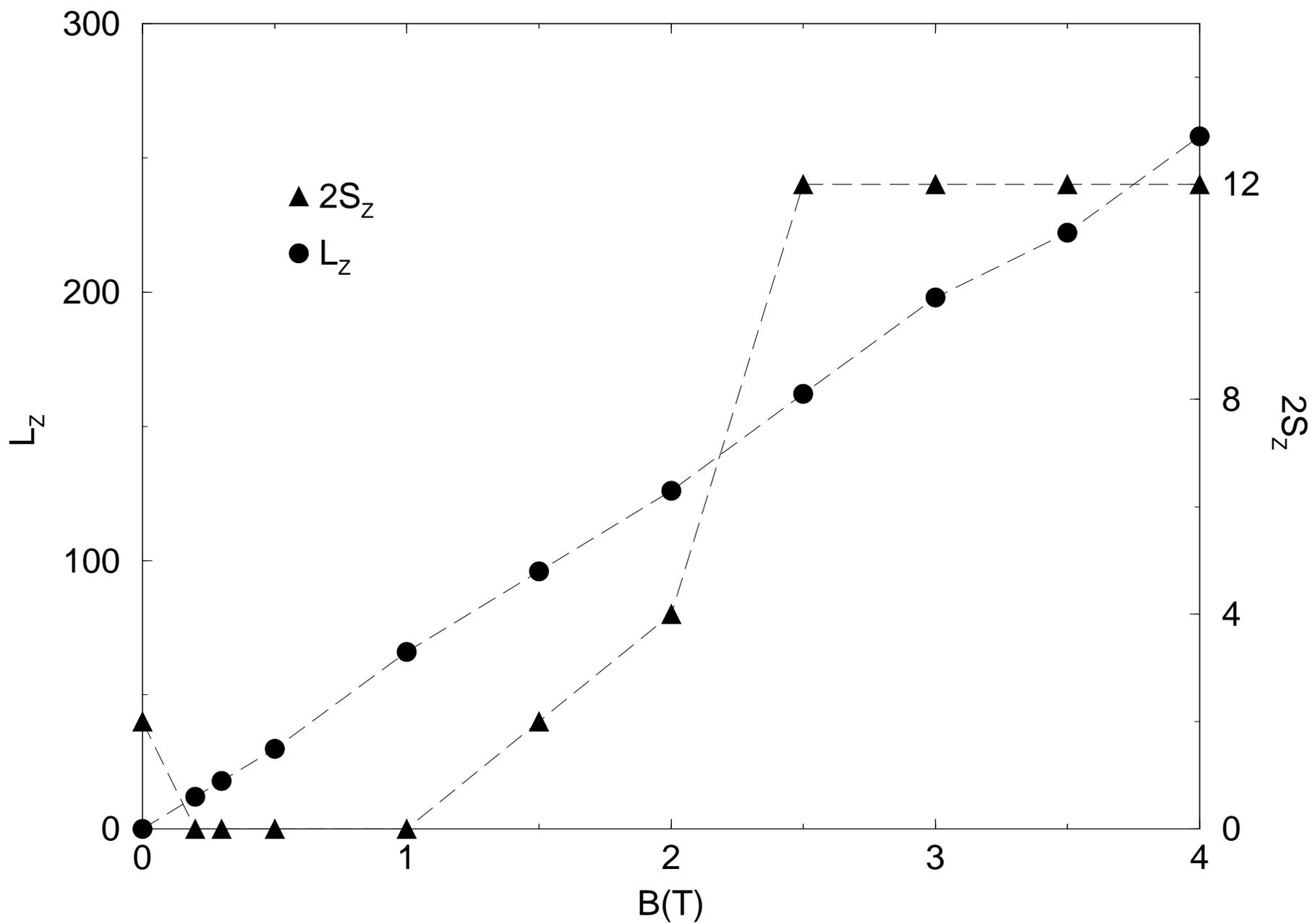
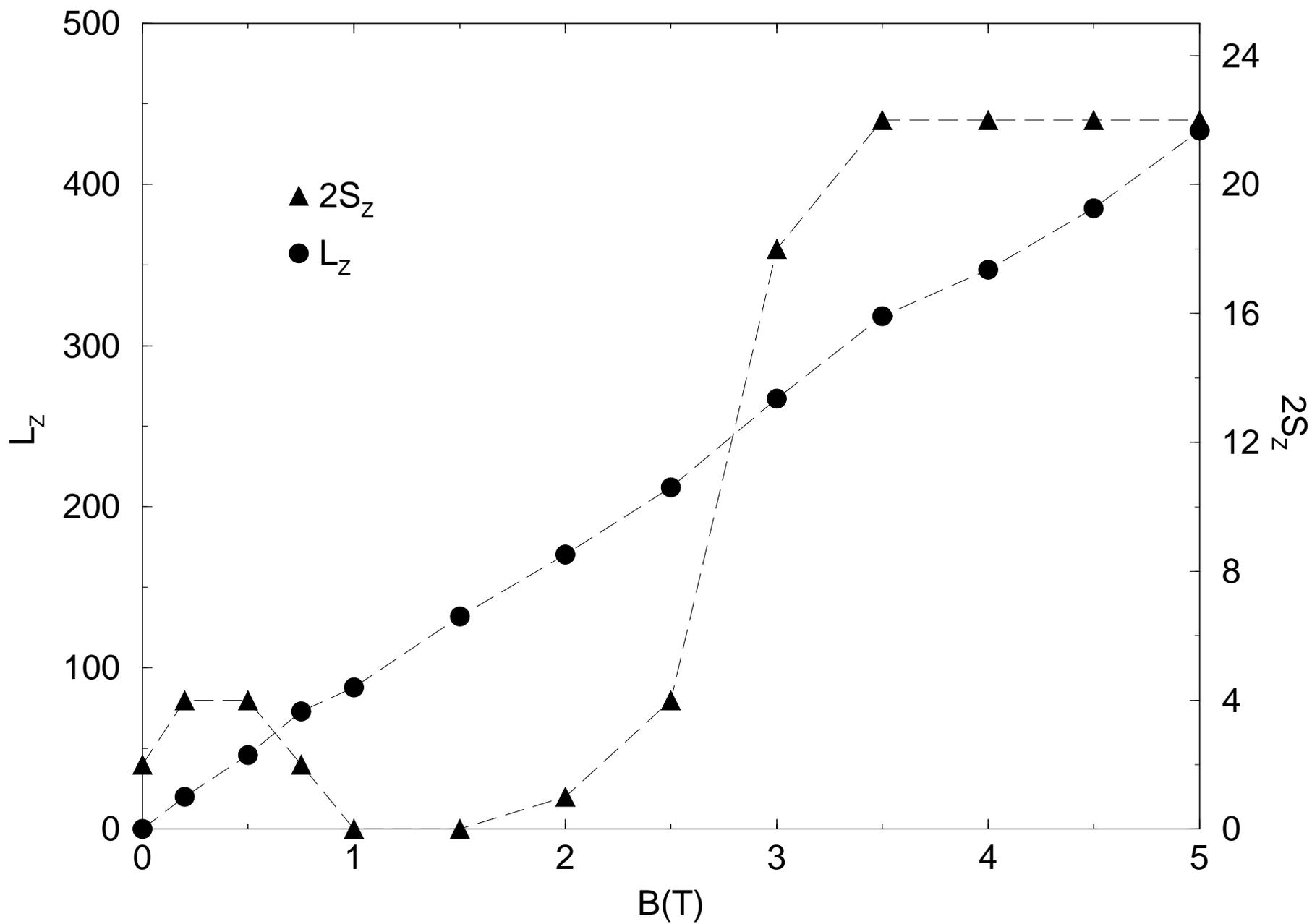
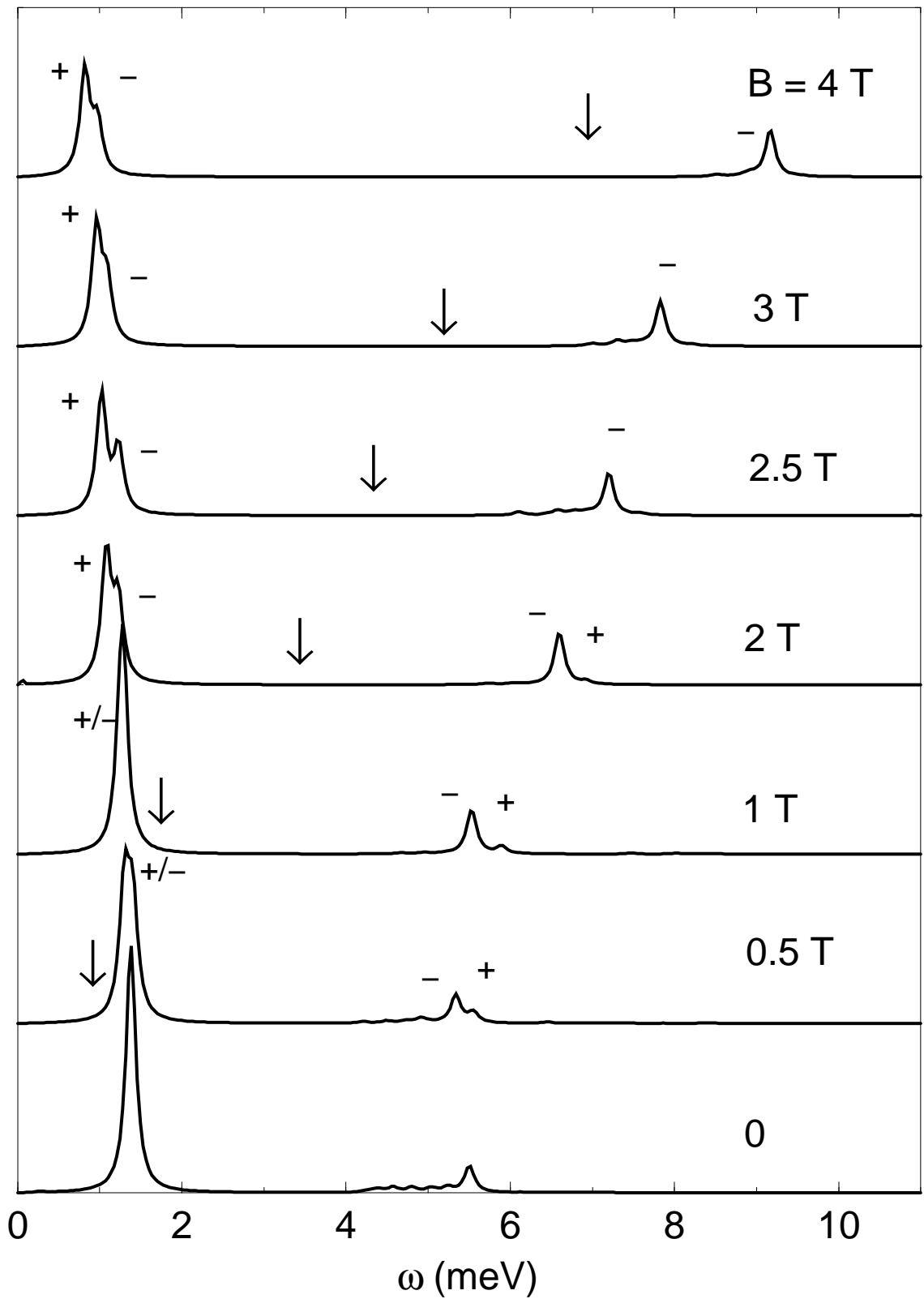
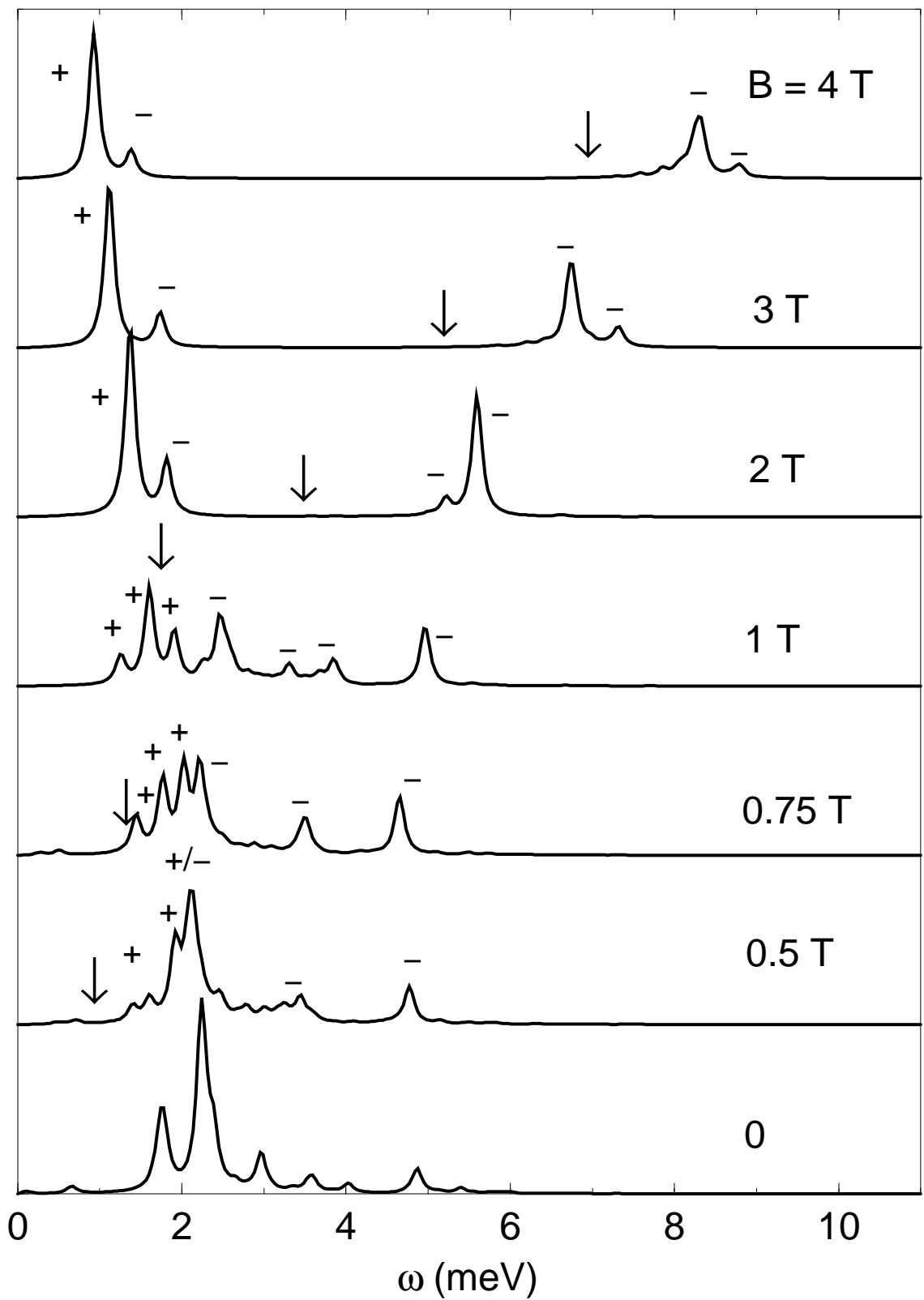
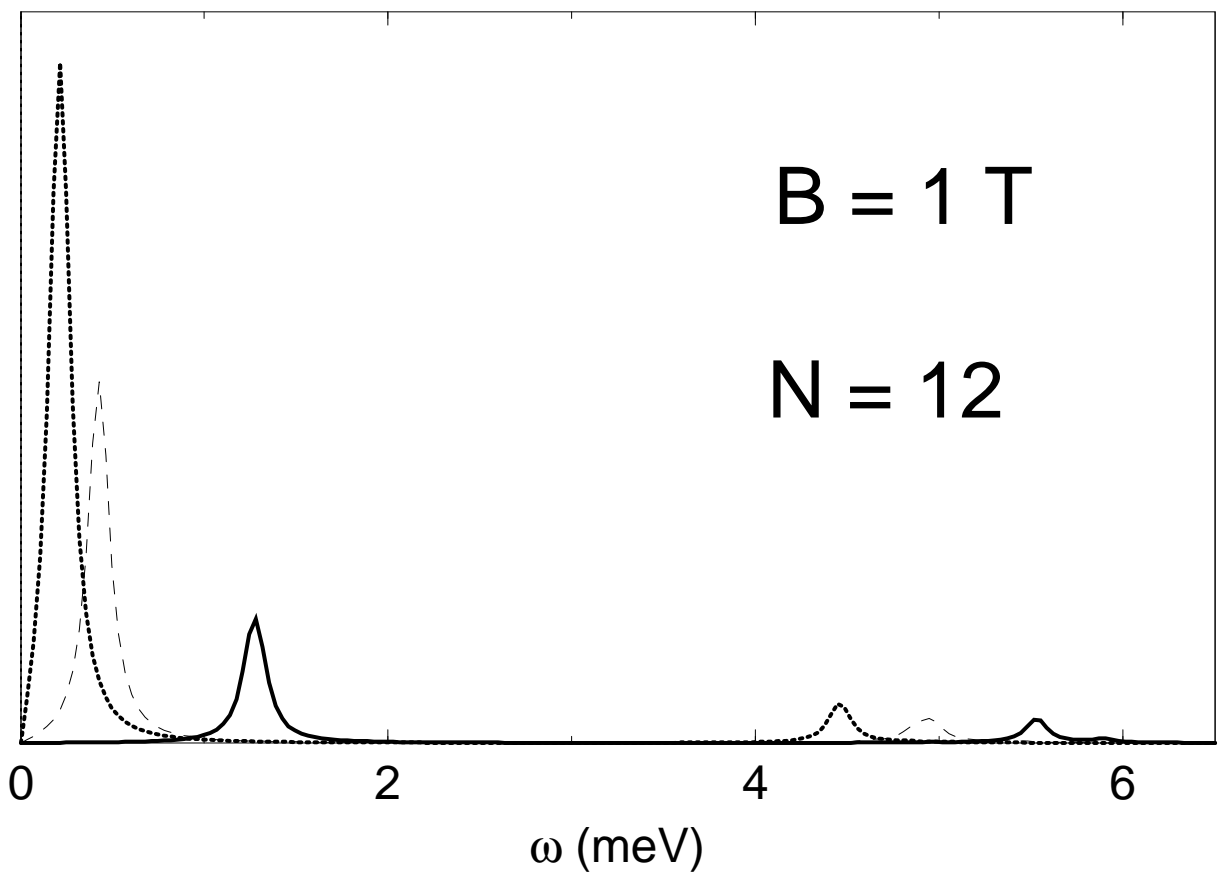
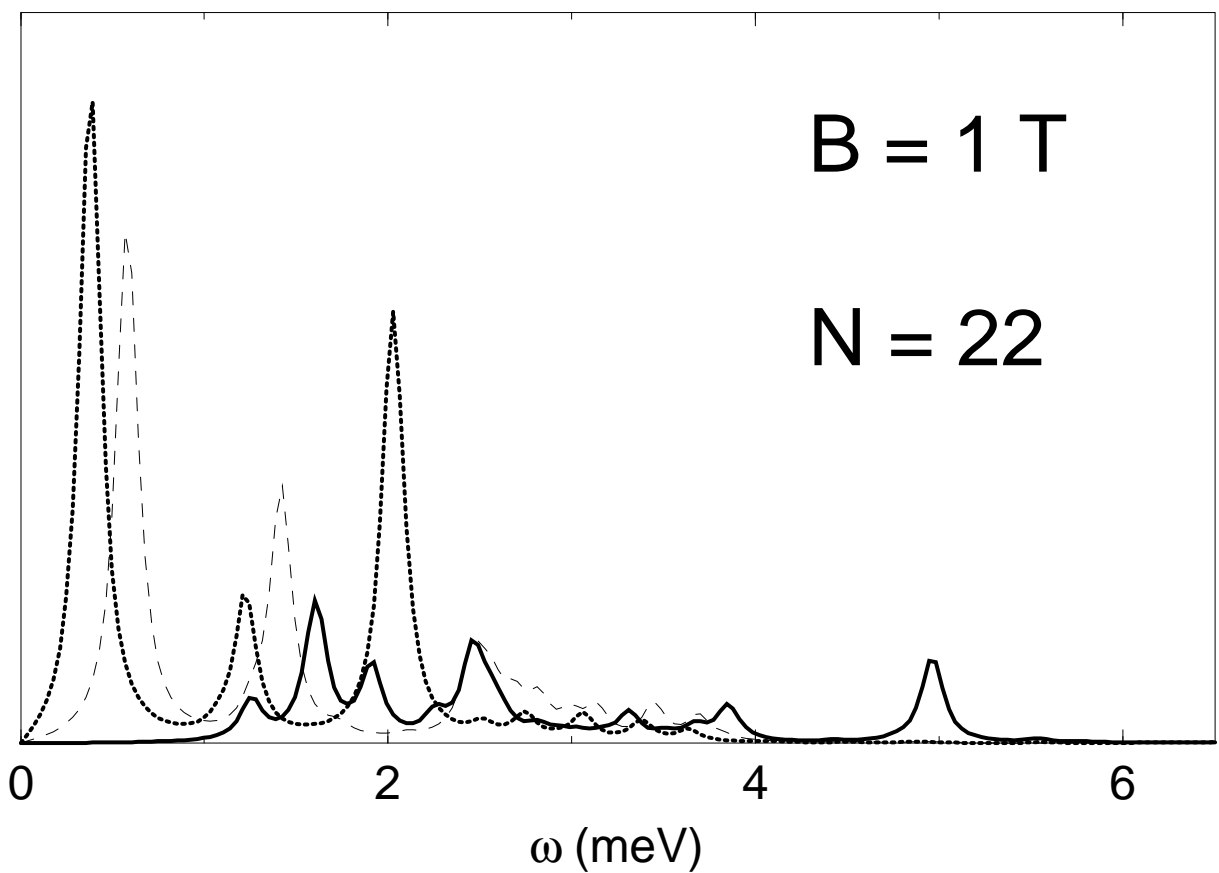
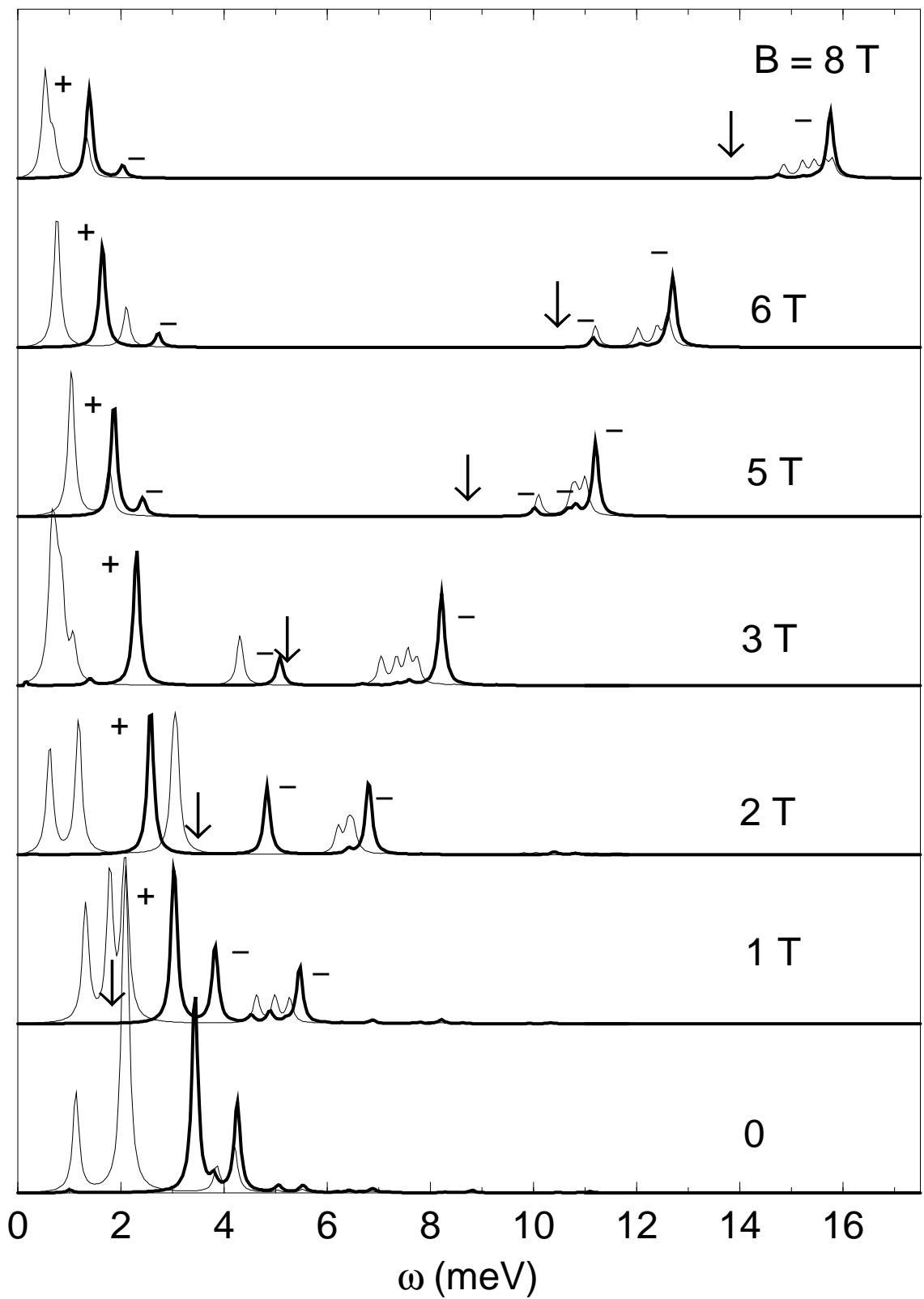
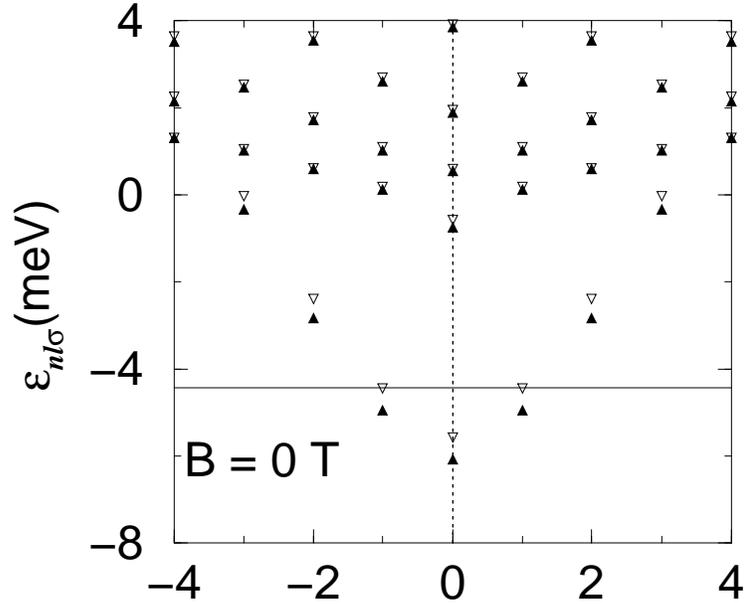
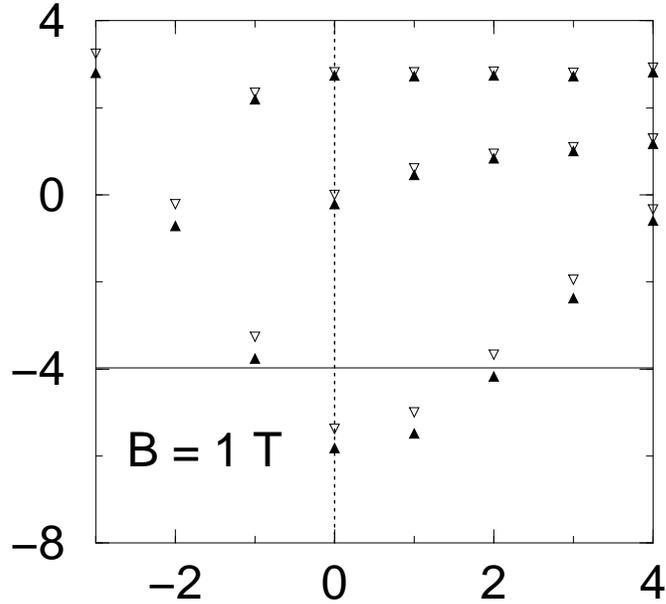
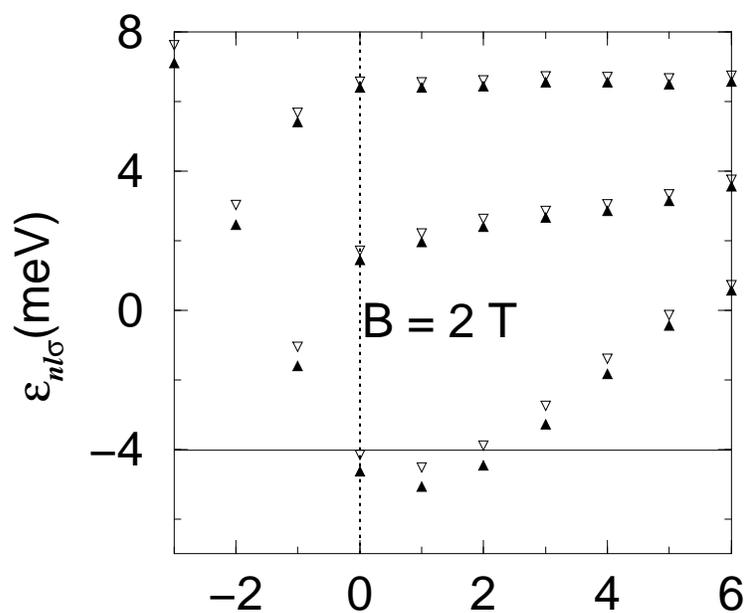
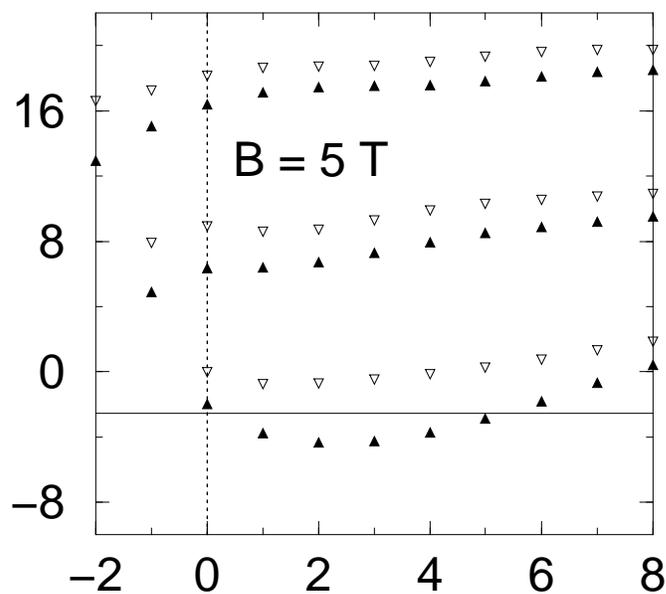
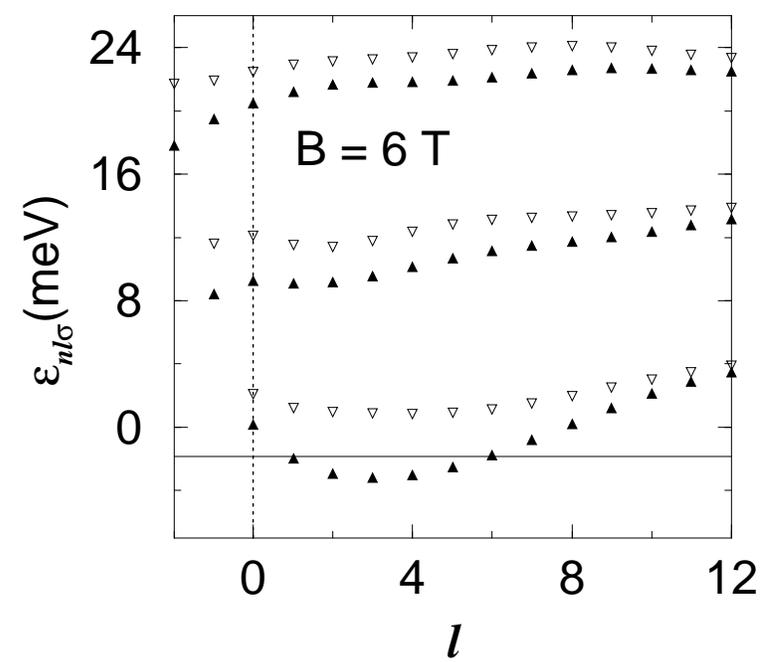
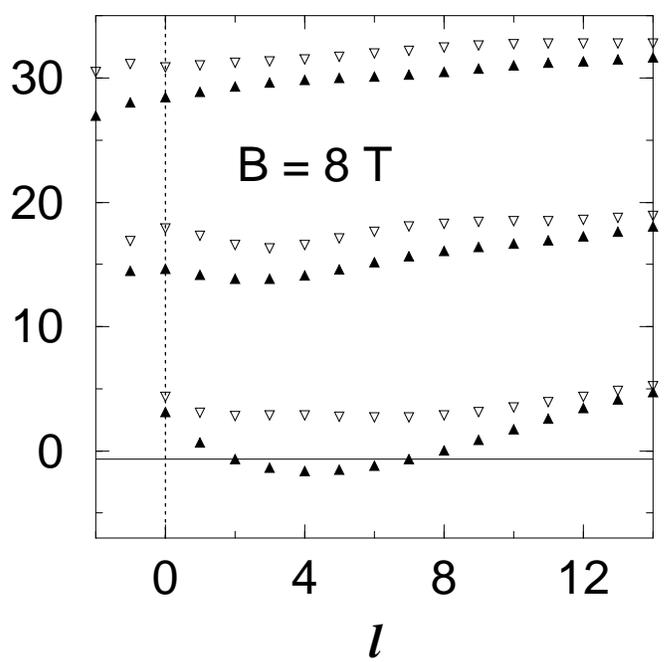
\end{document}